\documentclass{aa}  
\usepackage{graphicx}
\usepackage{longtable,lscape}
\usepackage[varg]{txfonts}

\usepackage{natbib}
\bibpunct{(}{)}{;}{a}{}{,}
\begin{document}
   \title{Molecular gas and a new young stellar cluster in the far outer Galaxy\thanks{Based on observations collected at the ESO 8.2-m VLT-UT1 Antu telescope (program 66.C-0015A).}
}


   \author{Jo\~ao L. Yun
		\and
		Davide Elia
		\and
          Pedro M. Palmeirim
	\and
	Joana I. Gomes
	\and
	Andr\'e M. Martins
          }


   \institute{Universidade de Lisboa - Faculdade de Ci\^encias \\
Centro de Astronomia e Astrof\'{\i}sica da Universidade de Lisboa, \\
Observat\'orio Astron\'omico de Lisboa, \\
Tapada da Ajuda, 1349-018 Lisboa, Portugal\\
             \email{yun@oal.ul.pt, eliad@oal.ul.pt, ppalm@oal.ul.pt, jgomes@oal.ul.pt, amartins@oal.ul.pt}
             }

   \date{Received November 20, 2008; accepted March 26, 2009.}

 
  \abstract
   {}
   {We investigate the star-formation ocurring in the region towards IRAS~07527-3446 in the molecular cloud [MAB97]250.63-3.63, in the far outer Galaxy. We report the discovery of a new young stellar cluster, and describe its properties and those of its parent molecular cloud.
}
   {Near-infrared $JHK_S$ images were obtained with VLT/ISAAC, and millimetre line CO spectra were obtained with the SEST telescope. VLA archive date were also used.}
   {The cloud and cluster are located at a distance of 10.3 kpc and a 
Galactocentric distance of 15.4 kpc, in the far outer Galaxy. Morphologically, IRAS~07527-3446 appears as a young embedded cluster of a few hundred stars seen towards the position of the IRAS source, extending for about 2--4~pc and exhibiting sub-clustering.  The cluster contains low and intermediate-mass young reddened stars,  a large fraction having cleared the inner regions of their circumstellar discs responsible for $(H-K_S)$ colour excess. The observations are compatible with a $\le$ 5~Myr cluster with variable spatial extinction of between $A_V=5$ and $A_V=11$.
Decomposition of CO emission in clumps, reveals a clump clearly associated with the cluster position, of mass $3.3 \times 10^3$~$M_\odot$.
Estimates of the slopes of the $K_S$-band luminosity function and of the star-formation efficiency yield values similar to those seen in nearby star-formation sites.
These findings reinforce previous results that the distant outer Galaxy continues to be active in the production of new and rich stellar clusters, 
with the physical conditions required for the formation of rich clusters continuing to be met in the very distant environment of the outer Galactic disc.
}
   {}


\titlerunning{Gas and new young stars in the far outer Galaxy }
\authorrunning{Yun, Elia, Palmeirim, Gomes \& Martins}

   \maketitle
%

\section{Introduction}

Stars form in molecular clouds across the Galaxy. Most molecular gas and dust exist in the Galactic disc and are most abundant toward the inner Galaxy \citep{clemens88}.
Thus, most star-formation sites, containing mostly young stellar clusters, have been found and studied in regions of the inner disk and in nearby molecular clouds \citep[e.g.,][]{tapia91, strom93, mccaughrean94, horner97, luhman98}.

Young stellar clusters are gravitationally bound systems of stars that formed
from the same parent cloud. They usually represent a statistically significant sample of stars with similar distances, and chemical compositions.
Quite frequently, they exhibit a relatively wide range of masses, often containing main-sequence massive stars as well as low-mass pre-main-sequence stars.
A main piece of information to understand the star-formation
process is the stellar mass distribution that results from the fragmentation of molecular clouds. Whether it is universal or depends on local environments, initial mass, or luminosity functions of a young cluster can be used to address important questions about the star-formation process \citep[e.g.,][]{pandey05}.
Young embedded clusters, in particular, unlike more evolved open clusters, which have lost significant numbers of low-mass stars due to a combination of cluster dynamics and Galactic tides \citep{vesperini97}, can be used to probe the stellar initial mass function ina more accurate \citep[e.g.,][]{muench00}, and its possible variation in space and time.

Because stellar mass is not a directly observable quantity, the luminosity function has been most extensively used to infer the underlying mass function \citep{luhman00,muench03}. For embedded clusters, the near-infrared luminosity function is most adequate as it can be obtained directly from photometric observations of the stellar cluster members, whose spectral energy distributions peak in the near or mid-infrared, without the assumption of any theoretical mass-luminosity-age relations and their corresponding uncertainties.

More attention has been devoted to star-formation sites toward the outer Galaxy and at large Galactocentric distances \citep[defined as $>$13.5 kpc; this number is based on the radial distribution of CO emission from the studies of][]{digel96,heyer98}. 
Extensive studies were performed of the distribution of CO, and of H$_2$, in star-formation sites in the outer Galaxy as well as a comparison of molecular clouds across the Galaxy \citep{wouterloot89,wouterloot90,brand95}. Based on observations in the millimetre and radio frequencies (basically molecular gas lines, maser emission, and ionized gas continuum emission), several star-formation sites in the far outer Galaxy were found  \citep[e.g.][]{fich84, wouterloot88, rudolph96}.
As for the stellar content of star-formation sites in the far outer Galaxy, discovery of isolated young stellar objects located at an estimated Galactocentric distance of about 15-19 kpc were reported by \citet{kobayashi00}. \citet{santos00} reported the discovery of two of the most distant galactic young stellar clusters (located at a distance of 10.2 kpc and a Galactocentric distance of 16.5 kpc) embedded in a molecular cloud containing a CS dense core.  Similarly, \citet{snell02} identified 11 stellar clusters in the second Galactic quadrant with Galactocentric distances of between 13.5 and 17.3 kpc. \citet{yun07} reported the discovery of an embedded cluster of an age of about 4 Myr at a Galactocentric distance of 15~kpc in the outer Galaxy.
\citet{brand07} found an embedded cluster of about 60 stars at a Galactocentric distance of 20~kpc. \citet{yasui08} studied the stellar content of two embedded clusters near the edge of the Galaxy.

However, there are still relatively few known resolved young embedded stellar clusters in the far outer Galaxy. This has prevented a systematic analysis of the star-formation properties in regions very far from the Galactic center. These regions have lower metallicities, lower gas densities and lower ambient pressures, which could affect the star-formation process resulting in different rates of cloud evolution. A different distribution of stellar masses in a cluster (different IMF) could emerge and consequently a different feedback mechanism from the stellar to the molecular component.

As part of our study of very distant young embedded clusters in the third Galactic quadrant, we report here the discovery of a new young stellar cluster embedded in a molecular cloud [MAB97]250.63-3.63 \citep{may97}, located far in the outer Galaxy, seen towards IRAS~07527-3446.
At a distance of about 10.3 kpc, it is one of the most distant young embedded clusters known to date.

Section~2 describes the observations and data reduction. In Sect.~3, we present and discuss the results. A summary is given in Sect.~4.


\section{Observations and data reduction}

\subsection{Near-infrared observations}

Near-infrared ($J$, $H$ and $K_S$) images were obtained on 2000 November 11 using the ESO Antu (VLT Unit 1) telescope equipped with the short-wavelength arm (Hawaii Rockwell) of the ISAAC instrument. The ISAAC camera \citep{moorwood98} contains a 1024 $\times$ 1024 pixel near-infrared array and was used at a plate scale of 0.147 arcsec/pixel resulting in a field of view of 2.5~$\times$~2.5 arcmin$^2$ on the sky.
For each filter, 6 dithered sky positions were observed. Series of 15 images with individual on-source integration time of 4 seconds were taken in the $J$-band. Similarly, series of 20 images, each of 3-second integration time, were obtained in the $H$ and $K_S$ bands. 

The images were reduced using a set of our own IRAF scripts to correct for 
bad pixels, subtract the sky foreground, and flatfield the images. Dome flats were used to correct for the pixel-to-pixel variations in the response. The selected images were 
then aligned, shifted, trimmed, and co-added to produce a final mosaic image for each band $JHK_S$. 
Correction for bad pixels was made while constructing the final mosaics, which cover about $3 \times 3$ arcmin$^2$ on the sky.
Due to the presence of significant field distortion in these ISAAC images (two pixels at the edges and 2.5 pixels in the corners), the individual tiles of mosaics were distortion corrected before they were tiled together. The correction for field distortion was performed using IRAF/{\tt geotran} together with the adequate correction files provided by ESO at their web page. 

The central, coadded region, with enhanced signal-to-noise ratio covers about 2.8~$\times$~2.8 arcmin$^2$ on the sky.
Point sources were extracted using {\tt daofind} with a detection threshold of 
4$\sigma$. The images were inspected to look for false detections that had been included by {\tt daofind} in the list of detected sources. These sources were eliminated from the source list and a few additional sources were added in by hand. Aperture 
photometry was made with a small aperture (radius = 2 pix, which is about
the measured FWHM of the point spread function) and aperture corrections, 
found from bright and isolated stars in each image, were used to correct 
for the flux lost in the wings of the PSF. The error in the determination 
of the aperture correction was $<$ 0.05 mag in all cases. 
A total of 630, 548, and 732 sources were found to have fluxes in $J$, 
$H$, and $K_S$, respectively, and errors $\sigma_{Ks} <$ 0.15 mag in the $K_S$ band.

The $JHK_S$ zero points were determined using faint infrared standard stars  \citep[S677-D and P545-C of the LCO/Palomar NICMOS Table of Photometric Standards; see][]{persson98} and checked with those obtained using 2MASS stars brighter than $K_S$ = 13.5 mag from the 2MASS All-Sky Release Point Source Catalogue \citep{skrutskie06,cutri03}.
Photometry errors are in the range from about 0.08 mag for the bright stars to 0.15 mag for the fainter ones. From the histograms of the $JHK_S$ magnitudes, we estimate the completeness limit of the observations to be roughly 20.0 magnitudes in the $J$-band, and 18.5 magnitudes in the $H$, and $K_S$ bands.

\subsection{CO mm line observations}
Millimetre single-dish observations of IRAS~07527-3446 were carried out at the 15~m 
SEST telescope (ESO, La Silla, Chile), during two observational campaigns, 
in 2001~May and 2002~December, respectively. 
Three maps were obtained in the rotational lines of $^{12}$CO(1$-$0) (115.271~GHz), 
$^{13}$CO(1$-$0) (110.201~GHz), and $^{13}$CO(2$-$1) (220.399~GHz), respectively. 
The SEST half-power beam width (HPBW) was $46\arcsec$ at 110~GHz 
and $22\arcsec$ at 230~GHz, so that, since the adopted grid spacing was 
$46\arcsec$, the $^{13}$CO(2$-$1) line emission was strongly undersampled. 
The $^{12}$CO(1$-$0) and $^{13}$CO(2$-$1) maps are composed of $10 \times 8$ pointings, 
while the $^{13}$CO(1$-$0) map is smaller ($6\times 7$ pointings).
These sizes correspond, on the sky, to an area markedly larger than 
the field of the VLT images, as discussed below.
Attempts were made to detect the emission from the C$^{18}$O(1$-$0), C$^{18}$O(2$-$1), CS(2$-$1), CS(3$-$2) transitions toward the central (IRAS source) position, using $t_{\mathrm{int}}=180$~s, but no spectral line was found.

The adopted integration time was $t_\mathrm{int}=60$~s for each pointing. 
A high-resolution 2000 channel acousto-optical spectrometer was used as a
back end, with a total bandwidth of 86~MHz and a channel width of 43~kHz; 
it was split into two halves to measure both the 115 and 230~GHz receivers 
simultaneously (in particular, the $^{12}$CO(1$-$0) and the $^{13}$CO(2$-$1) lines were observed simultaneously, as were the $^{13}$CO(1$-$0) and (2$-$1) lines).
At these frequencies, the aforementioned channel 
width corresponds to approximately 0.11~km~s$^{-1}$ and 0.055~km~s$^{-1}$, 
respectively. 
The spectra were taken in frequency-switching mode, 
recommended to save observational time when mapping extended sources.
The antenna temperature was calibrated with the standard chopper wheel method.
Pointing was checked regularly towards known circumstellar
SiO masers; pointing accuracy was estimated to be higher than
$5\arcsec$.

The data-reduction pipeline consisted of the following steps: 
\textit{i)}  folding the frequency-switched spectrum; 
\textit{ii)} fitting the baseline by a high-order polynomial and subtracting it; 
\textit{iii)} coadding repeated spectra obtained at the same sky position, weighting by both the integration time and the inverse of the system temperature; 
\textit{iv)} obtaining the main beam temperature $T_{\mathrm{MB}}$ by dividing the antenna 
temperature $T_{\mathrm{A}}$ by the $\eta_{\mathrm{MB}}$ factor, equal to 0.7 for 
the first two lines (at 110-115~GHz) and 0.5 for the third one (at 220~GHz); and
\textit{v)} finally, rebinning the spectra to obtain the same LSR velocity grid
for all the lines, with an exact channel separation
$\Delta V_{\mathrm{chann}} = 0.12$ and 0.06~km~s$^{-1}$ at 115 and 230~GHz, 
respectively.

The spectrum baseline RMS noise (in $T_{\mathrm{MB}}$), averaged over 
all map positions, has been found to be 0.2~K for $^{12}$CO(1$-$0) and
$^{13}$CO(1$-$0), and 0.15~K for $^{13}$CO(2$-$1).

The same data-reduction pipeline was applied to the spectra of 
the aforementioned transitions with no detection, in order to evaluate the values of the RMS noise to be adopted as a 1~$\sigma$ upper limit for those lines. The values obtained are 0.1~K for C$^{18}$O(1$-$0), 0.2~K for C$^{18}$O(2$-$1), 0.09~K for CS(2$-$1), and 0.07~K for CS(3-2).

\subsection{VLA centimetre observations}

We searched for continuum observations made with the Very Large Array (VLA) of the National Radio Astronomy Observatory (NRAO)\footnote{The National Radio Astronomy Observatory is a facility of the National Science Foundation operated under cooperative agreement by Associated Universities, Inc.} 
to look for a radio continuum counterpart to IRAS~07527-3446.
The VLA data archive contained continuum observations at 6~cm towards IRAS~07527-3446, which were carried out with the VLA in its C and D configurations on 1993 October 3. The observations were made in both left and right circular polarization, with an effective bandwidth of 100 MHz. The phase centre of the observations was $\alpha(2000)$ = 7h54m37.2s, $\delta (2000) =-34^{\circ} 54' 49''\!\!.0$. The data were reduced and calibrated using the Astronomical Image Processing System (AIPS) software of the NRAO. The flux calibrator was 3C286, with an adopted flux density of 7.5 Jy, while the phase calibrator was 0826-373, with a bootstrapped flux density of 2.3$\pm$0.01 Jy. The image had an rms sensitivity of 160 $\mu$Jy/beam, with a size (FWHM) and position angle (P.A.) of the synthesized beam of $18''\!\!.0\times 13''\!\!.1$ and $-23^{\circ}$, respectively.

\section{Results and discussion}

\subsection{The young stellar cluster: identification and morphology}

\begin{figure}
   \centering
   \includegraphics[width=9cm]{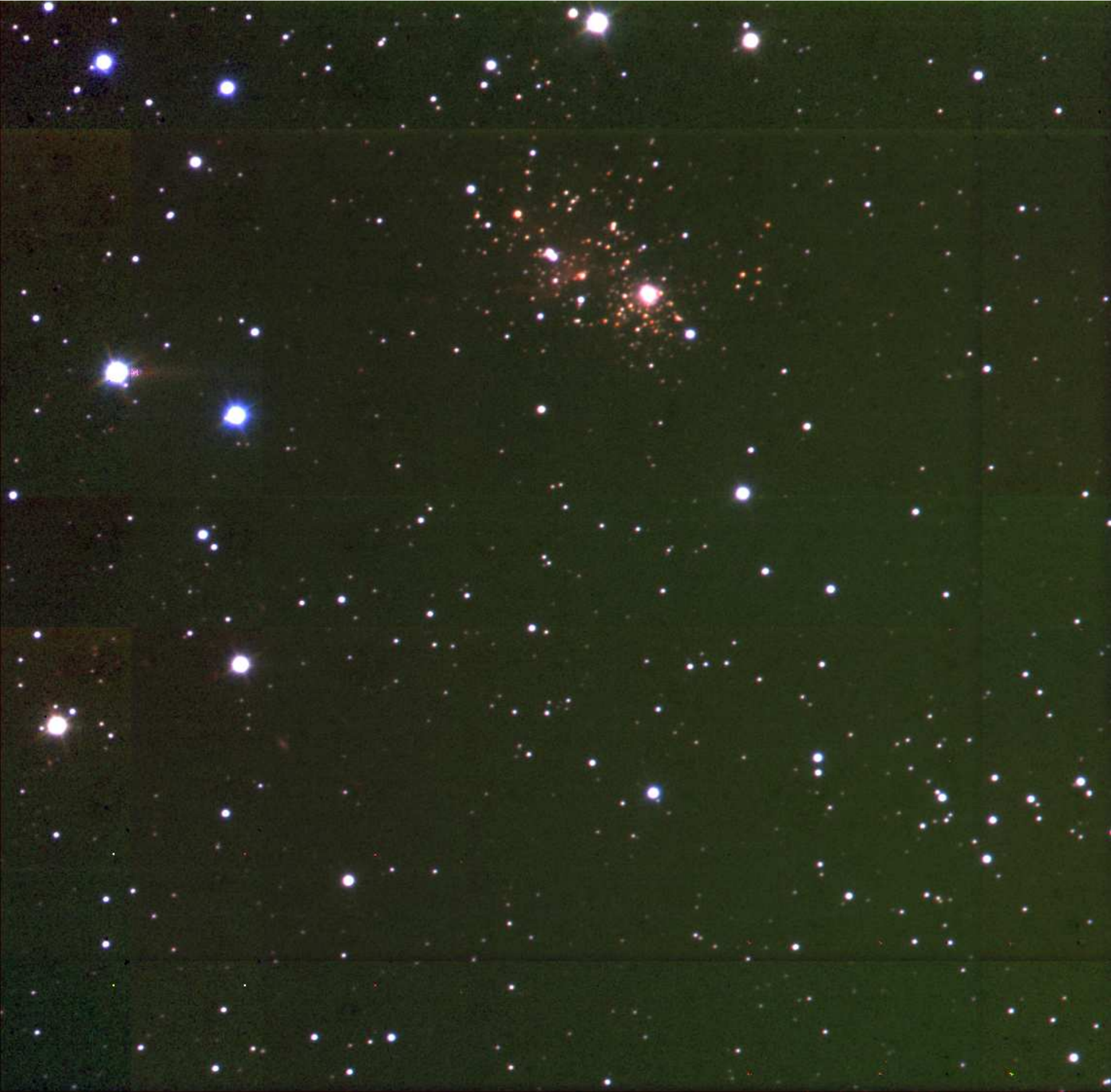}
   \caption{
$J$ (blue), $H$ (green), and $K_S$ (red) colour composite image towards IRAS~07527-3446. A new young red cluster is seen in the northern part of the image. The image covers about $2.8\times 2.8$ arcmin$^2$. North is up and East to the left.
	} 
	\label{JHK}%
    \end{figure}

Figure~\ref{JHK} presents the ISAAC $JHK_S$ colour composite image obtained towards IRAS~07527-3446. 
A reddened stellar cluster is clearly seen in the northern part of the image, where the presence of a high concentration of stars resides. 
In this final, mosaicked image, IRAS~07527-3446 (and the cluster it represents) is not located at the central position. This is the result of a deliberate choice for the position of the centre of the initial frames, about 1$^{\prime}$ south of the IRAS source nominal position. 
This choice was made to avoid contamination of the images by an extremely bright, unrelated star (HD 64876: $K_S=2.6$, $V=6.1$) located about 4$^{\prime}$ northward of the IRAS source.

\begin{figure}
   \centering
   \includegraphics[width=9cm]{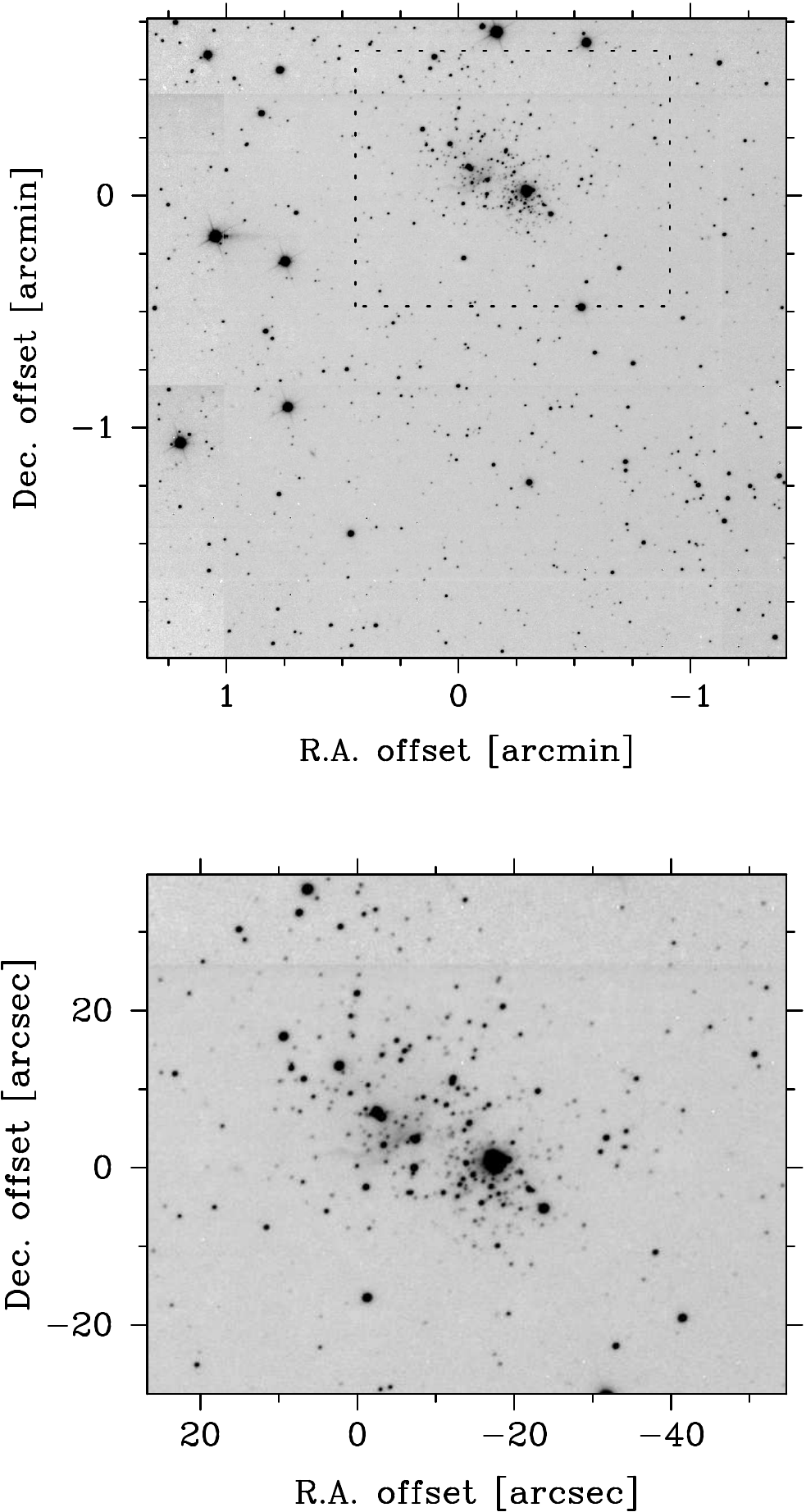}
   \caption{
$K_S$-band image towards IRAS~07527-3342. The dashed box indicates the ``cluster region (see Sect. 3.3.1)'' seen in more detail in the lower panel (close-up view of $81'' \times 66''$ containing the cluster). The axes give the coordinates relative to the IRAS source ($\alpha(2000)$ = 7h54m37.2s, $\delta (2000) =-34^{\circ} 54' 49''\!\!.0$). North is up and East to the left.
	} 
	\label{K}%
    \end{figure}

A simple visual inspection of Fig.~\ref{JHK} reveals that the stars appear to form two distinct groups or sub-clusters: a northeastern (NE) group and a southwestern (SW) group. 
The SW group appears to be richer with a more significant central concentration of stars where the brightest star is located. The NE group appears less extended and more sparse but also redder. In addition, the NE group includes the presence of tenuous diffuse extended emission. This can be seen more clearly in Fig~\ref{K}. The lower panel of this figure shows what we define below as the ``cluster region''. A careful analysis of this panel reveals the presence of faint diffuse extended (nebular) emission, seen mostly towards the NE group.

%

%

\begin{figure}
   \centering
   \includegraphics[width=9cm]{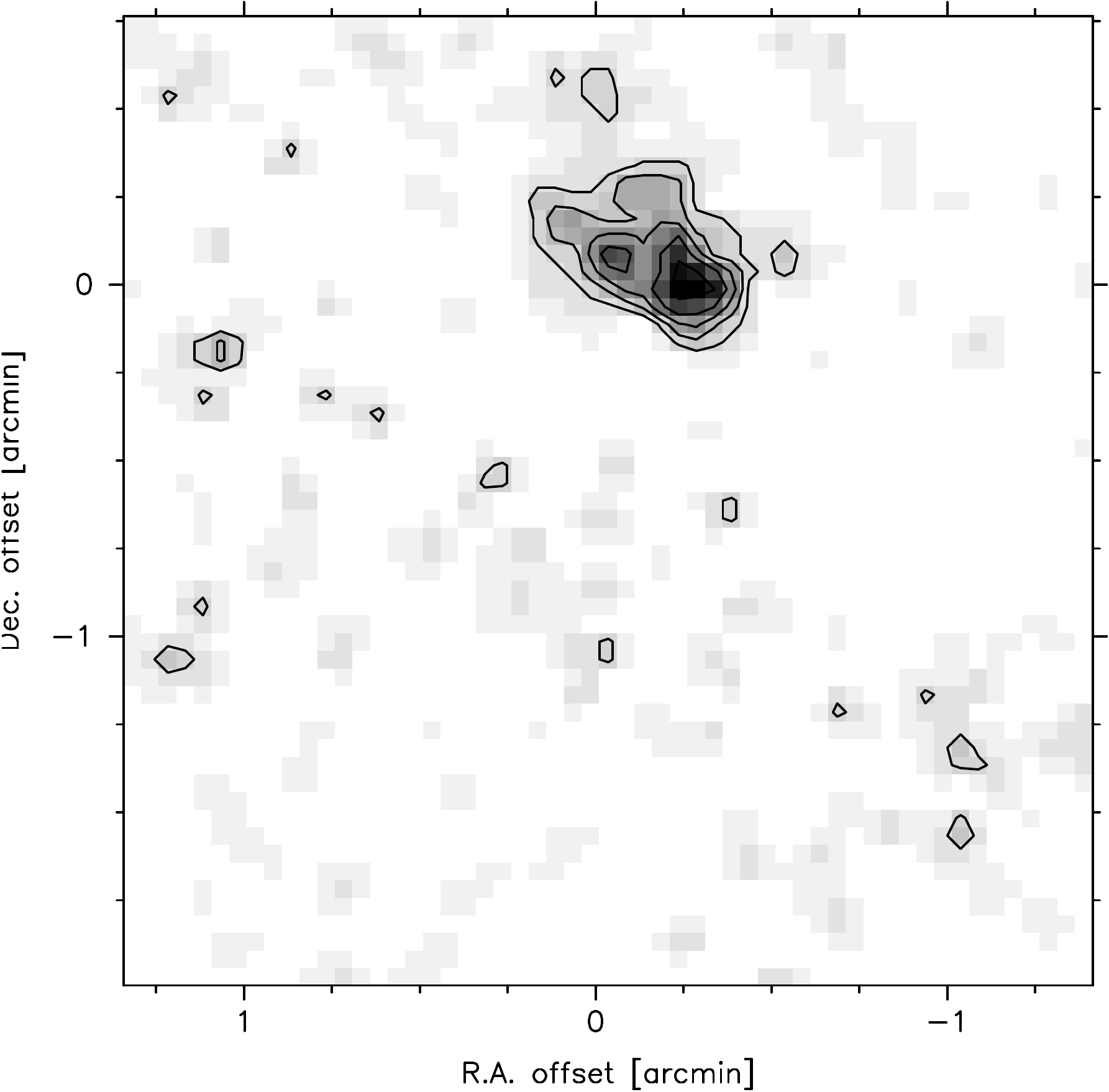}
   \caption{
Contour plot of the stellar surface density towards IRAS~07257-3446. The contour 
levels are 3, 6, 10, 15, and 22 $\sigma$ above the background mean stellar density. 
A main stellar cluster is easily identified with evidence for sub-clustering. Offsets are relative to the IRAS source ($\sigma=40$ stars pc$^{-2}$, for a distance of 10.3 kpc).
	} 
	\label{density}%
    \end{figure}

In order to detect the presence of the cluster and any possible sub-clustering in a more robust way, a contour plot of the stellar surface density was made and is shown in Fig.~\ref{density}.
The contour levels are set to show significant enhancements above the average field star density. In this figure, a clear concentration in the spatial distribution of stars is seen, which is identified as a stellar cluster. The contours marking this cluster are elongated in a direction close to SW-NE (position angle of about 55$^\circ$). The concentration of stars at the strongest peak of the surface stellar density of the image (corresponding to the SW group) is very high ($\sim$~900 stars pc$^{-2}$; for a distance of 10.3 kpc, see below), about 60 times higher than the average field star density (which itself was determined by counting stars in the southern part of the image where the cluster is not located). Away from this peak, in the transversal direction (perpendicular to the elongation), the cluster merges with the field for distances of the order of 16$^{\prime\prime}$ ($\sim$0.8 pc, see below).
The cluster appears to occupy an area, which when approximated by an ellipse, has major and minor axes of 2.3 and 1.7 pc, respectively.
Down to the $K_S$-band magnitude limit of 18.5, the mean surface density in this cluster (after subtraction of the average field star density of 14 pc$^{-2}$) 
is 45 stars pc$^{-2}$. Within the ellipse defined above, the expected number of 
cluster members is $\sim$~138 stars, and the expected number of field stars is $\sim$~44 stars.

A second peak is also clearly present in the surface stellar density map. It corresponds to the NE group of stars and its peak value is two-thirds of the stronger SW peak.

%

\subsection{The molecular cloud}

\subsubsection{Molecular gas and distance}

The IRAS~07527-3446 position falls within the centroid of a CO cloud listed in \citet{may97}, of their catalog of southern outer Galaxy molecular 
clouds  (for this cloud, hereafter we adopt the  SIMBAD nomenclature [MAB97]250.63-3.63).
Our CO maps presented here represent a portion of this cloud, observed with unprecedented resolution, considering that the sampling interval of \citet{may97}
was $7\arcmin.5$.  
The CO maps are significantly larger than the field imaged in the NIR, giving information about the gas distribution in that zone on a large scale, the cluster itself 
being covered only by the (0,0) and (1,0) pointings.

Only lines characterized by a signal-to-noise ratio \textit{SNR}~$>3$ were 
considered as genuine emission. Out of the 80 observed points, significant 
$^{12}$CO(1$-$0) and $^{13}$CO(2$-$1) emission was detected from 43 and 26 of 
them, respectively, while for $^{13}$CO(1$-$0) emission was detected from 18 
pointings out of 42.

\begin{figure}
\centering
 \includegraphics[width=9cm]{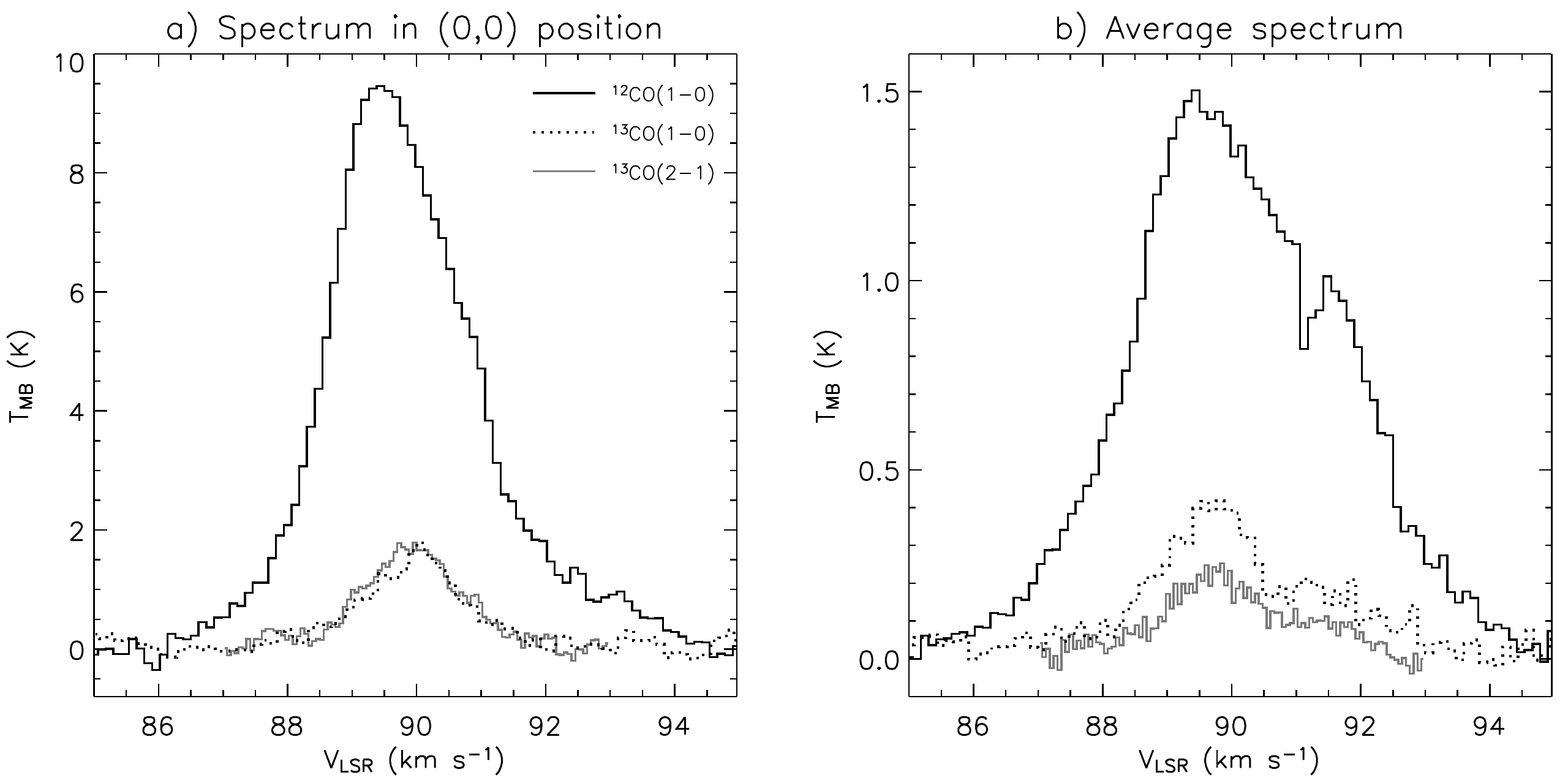}
\caption{Panel $a$: spectra of the three observed lines towards the IRAS source ((0,0) position of the maps). Panel $b$: Average of all spectra in the map, for each line. 
}\label{spectra}
\end{figure}

Figure~\ref{spectra} (panel~$a$) shows the spectra taken towards IRAS~07527-3446  ((0,0) position of the map). The $^{12}$CO(1$-$0) spectrum peaks at about 89~km~s$^{-1}$, while those of 
both $^{13}$CO lines peak at about 90~km~s$^{-1}$; these values can be adopted as the velocity of the gas component associated with the cluster, and are consistent with the global radial velocity of [MAB97]250.63-3.63 quoted by \citet{may97}.
The bulk of the CO emission is in the range $V_{\mathrm{lsr}} \simeq
85 - 95$~km~s$^{-1}$ for the $^{12}$CO(1$-$0) line, and 
$V_{\mathrm{lsr}} \simeq 88 - 93$~km~s$^{-1}$ for the other two 
lines (Fig.~\ref{spectra}, panel $b$). 

The first piece of information that can be extracted from the spectra is the kinematic heliocentric distance, derived from the peak radial velocity. This can be done applying the circular rotation model by \citet{brand93}.
Using $V_{\mathrm{LSR}}=90$~km$^{-1}$, a distance $d\simeq 10.3$~kpc is obtained.
Given the Galactic longitude, this implies a Galactocentric distance of 15.4 kpc to the cloud, well into the far outer Galaxy and among the largest for known Galactic star-formation sites. We estimate a distance uncertainty of up to 20\% due to uncertainties in the rotation curve and possible streaming motions. 

\subsubsection{Molecular gas distribution}

\begin{figure}
\centering
 \includegraphics[width=9cm]{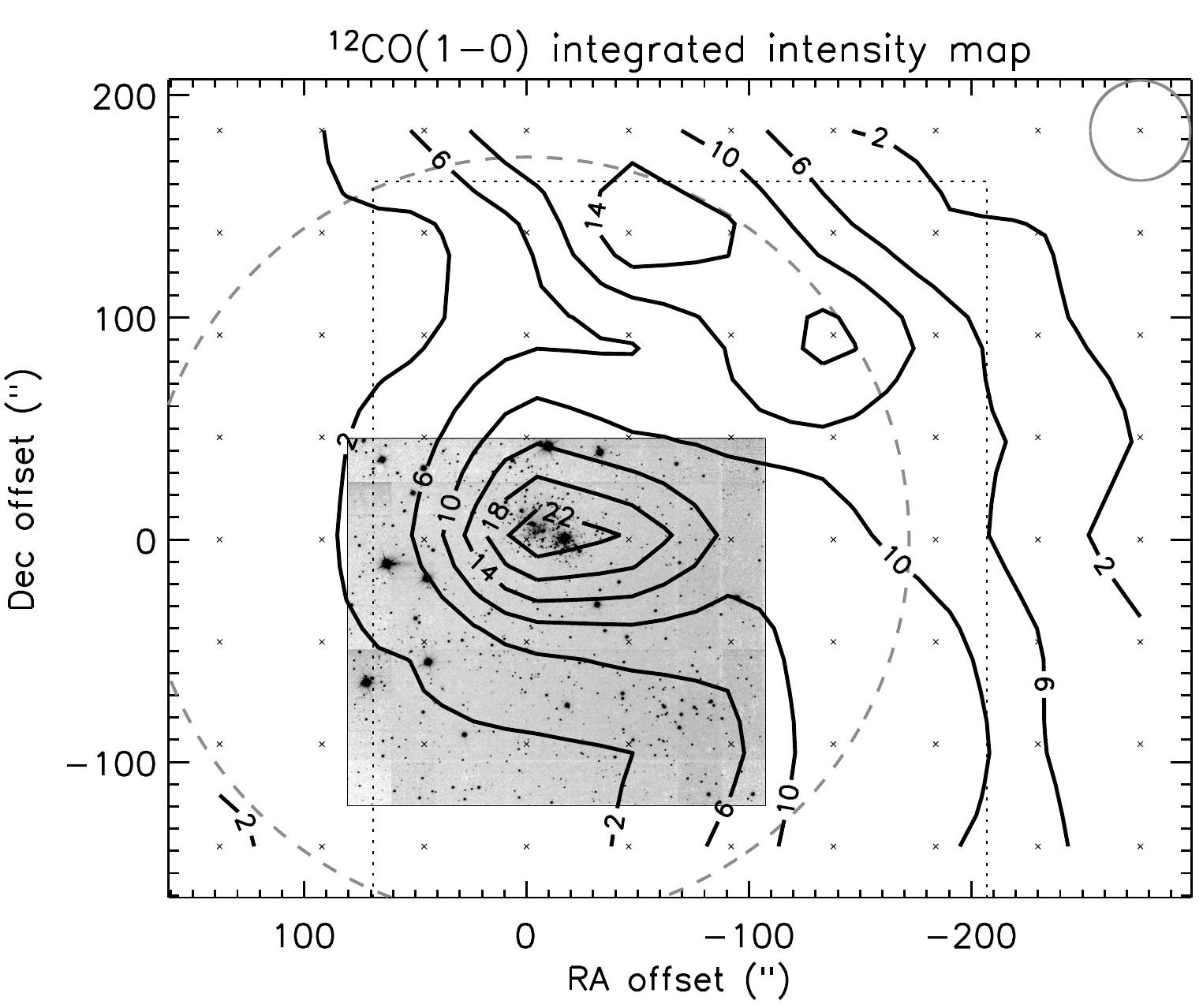}
\caption{Smoothed contours of $^{12}$CO(1$-$0) integrated intensity (in the velocity range between 84 and 96 km s$^{-1}$)
observed in the IRAS~07527-3446 region, superimposed on the $K$-band 
image of Fig.~\ref{K}. The lowest contour is at 2 K km s$^{-1} (8\sigma)$. Subsequent contours are in steps of 4 K km s$^{-1}$. 
The SEST beam at 115 GHz is displayed in the top right corner.
The dotted box delimits the area observed in $^{13}$CO(1$-$0).
The grey dashed circle represents approximatively the position and 
the average size of the gas clump (centred on the cluster position) obtained 
by decomposing the cloud with CLUMPFIND (see text). The small crosses indicate the observed positions.
}
\label{map12}
\end{figure}

\begin{figure}
\centering
\includegraphics[width=9cm]{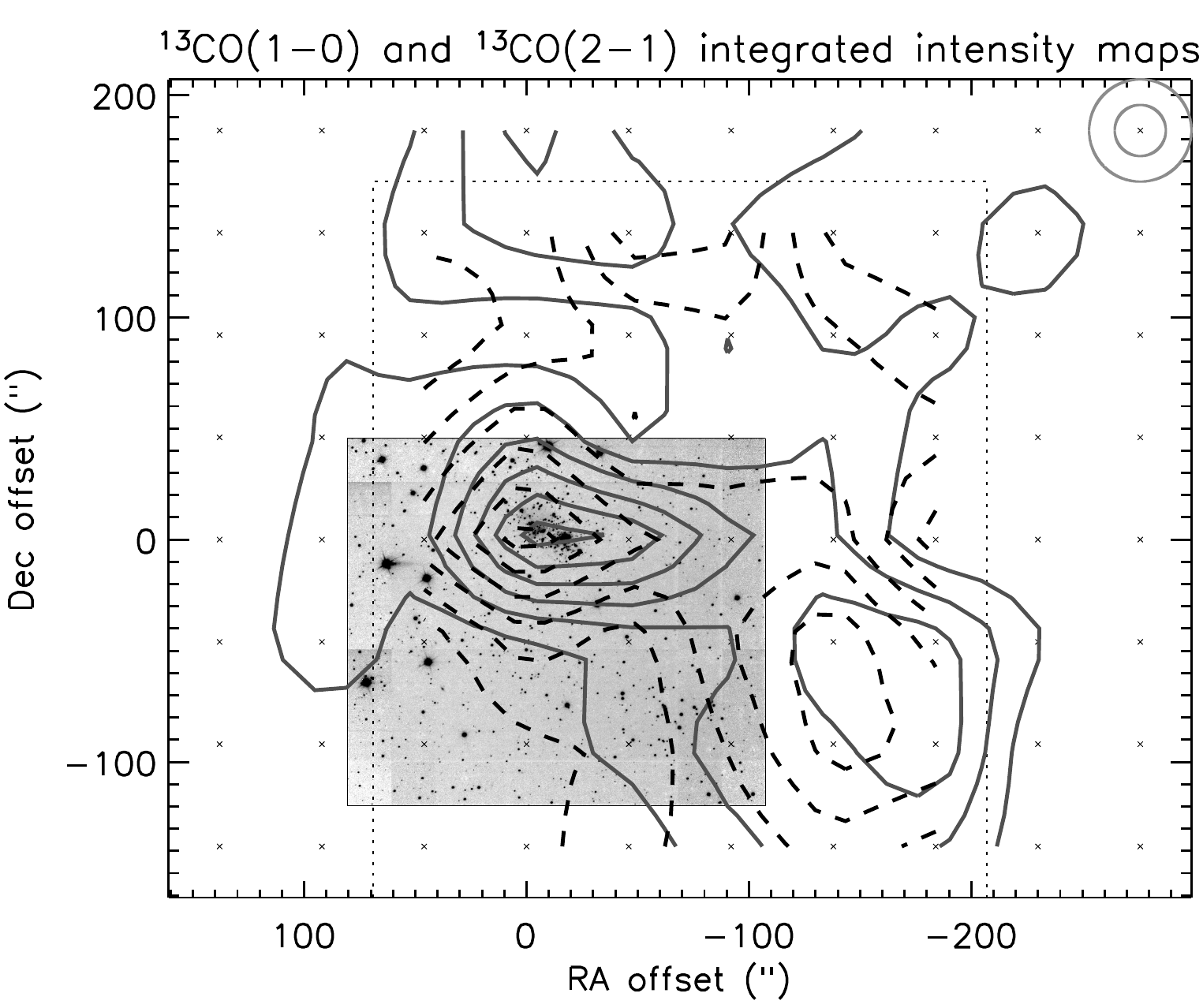}
\caption{As in Figure~\ref{map12}, but for $^{13}$CO(1$-$0) (dashed line) 
and $^{13}$CO(2$-$1) (grey solid line) transitions. In both cases, levels 
start at 0.5~K~km~s$^{-1}$ (2.5$\sigma$ for $^{13}$CO(1$-$0) and 5.5$\sigma$ for $^{13}$CO(2$-$1)) and are in steps of 0.5~K~km~s$^{-1}$; 
they are not labeled for the sake of clarity. The range of velocities are between 85 and 95 km s$^{-1}$ for $^{13}$CO(1$-$0) and between 87 and 93 km s$^{-1}$ for $^{13}$CO(2$-$1).
The SEST beams at 110 and 220~GHz are displayed in the top right 
corner by means of a large and a small grey circle, respectively. The dotted box delimits the area observed in $^{13}$CO(1$-$0).
}\label{map13}
\end{figure}

The contour map of the $^{12}$CO(1$-$0) integrated intensity $I=\int T_{\mathrm{MB}}~dv$ 
is shown in Fig.~\ref{map12}, superimposed on the $K$-band image. The $^{13}$CO(1$-$0) and 
$^{13}$CO(2$-$1) integrated intensity contour maps are shown in 
Fig.~\ref{map13}.

The morphology of the emission is very similar in all 
the maps, with the main (strongest) peak located at the cluster location.
The cluster appears to correspond to a strong gas concentration 
which emerges from a more diffuse filament. In more peripheral zones 
of the maps, it is possible to recognize the presence of more 
condensations, one north of the cluster, and another southwest of it, 
as highlighted by figures~\ref{map12} and \ref{map13}, and discussed in detail below.

\begin{figure}
\centering
\includegraphics[width=9cm]{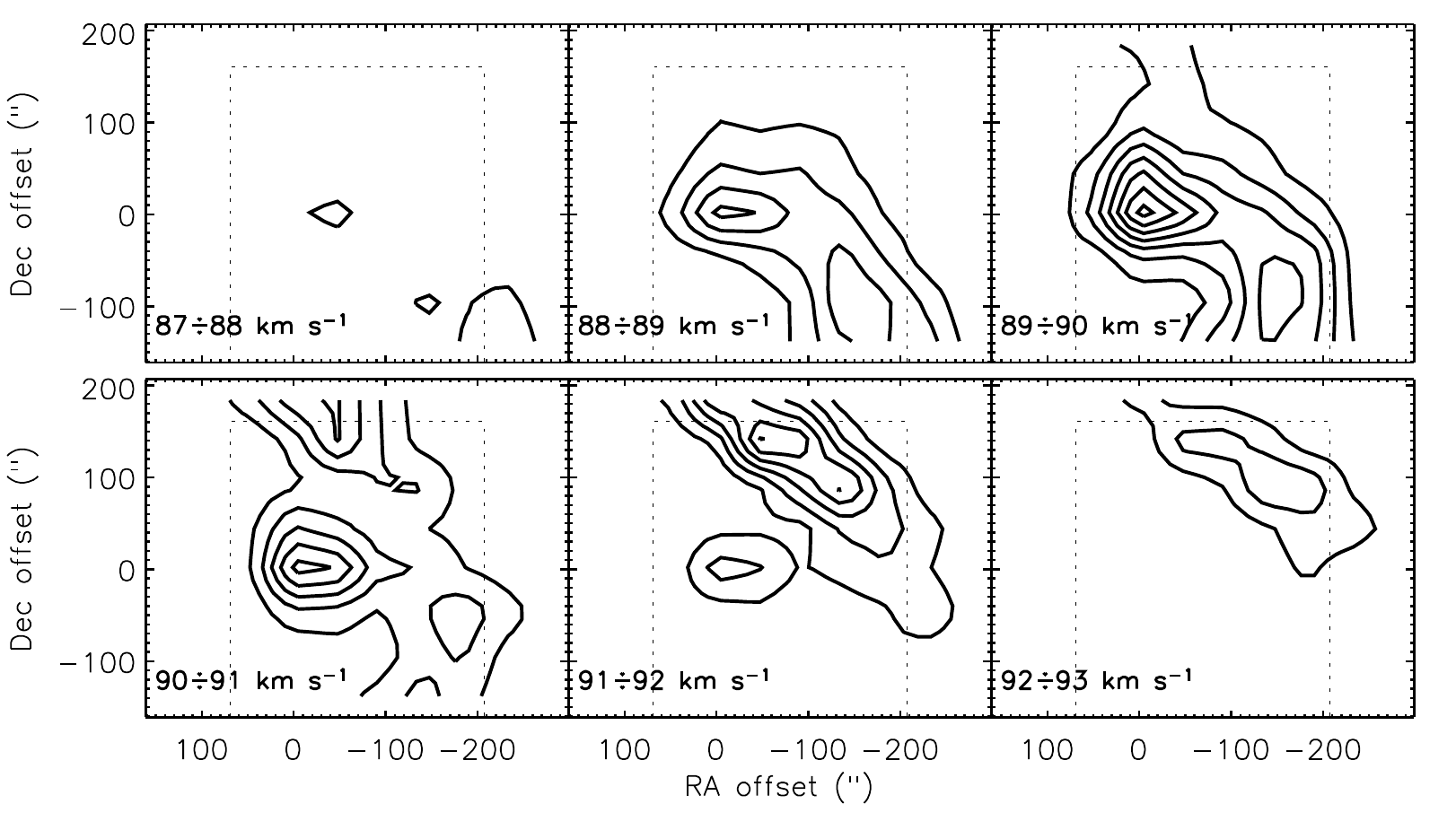}
\caption{Channel maps obtained integrating the $^{12}$CO(1$-$0)
emission over velocity ranges of width 1 km s$^{-1}$.
Contours are in steps of 1~K~km~s$^{-1}$. Dashed line 
delimits the area observed also in $^{13}$CO(1$-$0).}\label{ch12}
\end{figure}

\begin{figure}
\centering
\includegraphics[width=9cm]{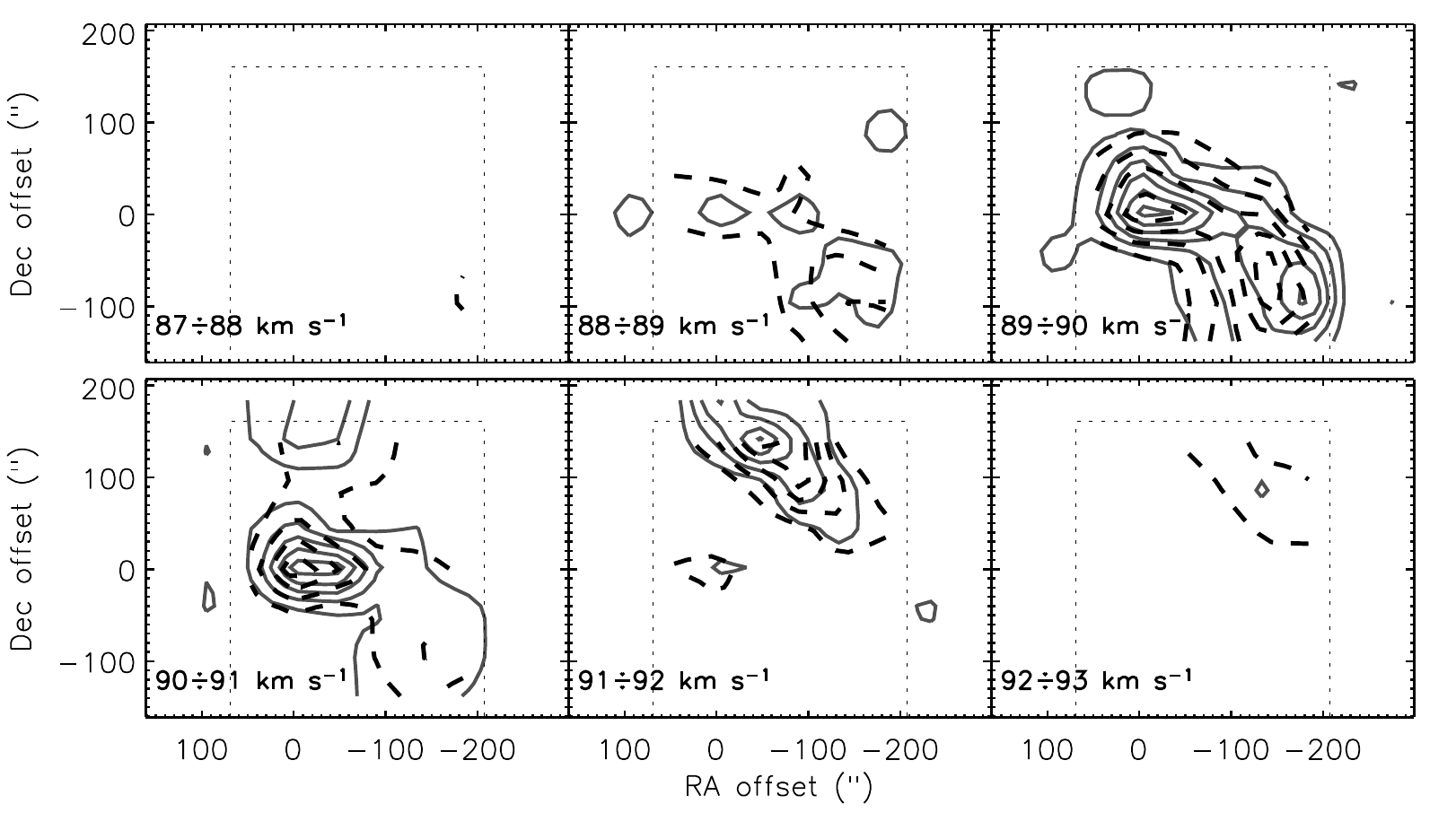}
\caption{As in figure \ref{ch12}, but for $^{13}$CO(1$-$0)(dashed line) 
and $^{13}$CO(2$-$1) (grey solid line).
Contours are in steps of 0.2~K~km~s$^{-1}$.}\label{ch13}
\end{figure}

Channel (velocity) maps were built from the line observations (figures \ref{ch12} 
and \ref{ch13}), where these ``clumps'' are seen to 
correspond to different velocity components. 
The clump in the northern 
part of the map is responsible for the second peak (at $V_{\rm LSR}=91.5$~km~s$^{-1}$) in the average
spectrum in Fig.~\ref{spectra}, panel~$b$). In terms of radial 
velocity, the central condensation corresponding to the cluster exhibits an intermediate velocity position.

\subsubsection{Molecular cloud mass and optical extinction}

A few different methods can be exploited to obtain
an estimate of the cloud mass from the line 
emission of molecular tracers, in this case the CO isotopes.
A first estimate can be made using the empirical 
correlation between the $^{12}$CO(1$-$0) integrated intensity and
the H$_{2}$ column density along the line of sight, $N_{i,j}$, at position $(i,j)$:
\begin{equation}\label{nh2}
N_{i,j}(\mathrm{H}_2)=X \int T_{\mathrm{MB}}~dv \qquad (\mathrm{cm^{-2}})\;,
\end{equation} 
where the constant $X$ has been determined empirically. 
Several authors quote different values for $X$, and it is accepted that, 
in general, it should vary with the Galactic position. 
In the southern outer Galaxy, including the
[MAB97]250.63-3.63 cloud, \citet{may97} used a value
of $X=3.8 \times 10^{20}$~cm$^{-2}$~(K~km~s$^{-1}$)$^{-1}$, which is almost
twice the values typically adopted for nearer clouds. Indeed,
\citet{strong96} derived a revised value for X, of
$1.9 \times 10^{20}$~cm$^{-2}$~(K~km~s$^{-1}$)$^{-1}$; it should be corrected according to
\citet{brand95}, who argued that in the far outer Galaxy it may be 30-45\% higher than
the local/inner Galaxy value. Thus, in this case we adopt
$X=2.8 \times 10^{20}$~cm$^{-2}$~(K~km~s$^{-1}$)$^{-1}$. To obtain the mass from
the column density, we also used a value of the mean molecular mass $\mu$
corrected for the gradient in He abundance along the Galactic radial
direction; at a Galactocentric distance of $R=17$~kpc, a correction of 1.28
on the mass value has to be applied to take in account this variation, and consequently $\mu$ varies from his 
``typical'' value of 2.72 \citep{all73} to 2.56 there \citep{wilson92}. Using
these corrected values, we obtain a mass of $M_{\mathrm CO}=1.3 \times
10^4 M_{\odot}$ for the mapped area.

This estimate is in good agreement with the upper limit represented by the
value of $3.2 \times
10^4 M_{\odot}$ obtained by \citet{may97} for the [MAB97]250.63-3.63
total mass using the same method. Scaled to their values of X and $\mu$, we find 
$1.7 \times 10^4 M_{\odot}$.

Furthermore, the derived $N_{0,0}(\mathrm{H}_2)$ value, of the column density towards the map reference position, can be used to calculate the optical extinction $A_V$ \citep[see][]{bohlin78}. Using the relation
$N(\mathrm{H}_2)/A_V=9.4 \times 10^{20}$~cm$^{-2}$~mag$^{-1}$ \citep{fre82}, we obtain 
$A_V=7.8$~mag, compatible with the value derived from infrared colour excesses (Section 3.2.2).

Another method for estimating the cloud mass consists of assuming LTE conditions and using 
both an optically thick line and an optically thin line, in this case $^{12}$CO(1$-$0), 
and $^{13}$CO(1$-$0) or $^{13}$CO(2$-$1), respectively. In this 
method \citep[see e.g.,][for details]{pin08}, the peak main beam 
temperature of the $^{12}$CO(1$-$0) line is used to derive its excitation temperature, which, for example, is found to be 12.5~K at the map reference
position. Then, assuming that excitation temperatures are the same 
for both lines, the optical depth and the column density of $^{13}$CO are calculated for each line of sight 
\citep[see also][for the specific case of $^{13}$CO(2$-$1)]{eli07}. 
Finally, to calculate the total mass, it is necessary to adopt a H$_2$/$^{13}$CO
abundance ratio, which in the far outer is expected to be greater than
the ``typical'' value of $5 \times 10^5$ quoted by \citet{dic78}; \citet{brand95} indeed found that for $R=17$~kpc, a ratio of $3 \times 10^6$
should be used.

The $^{13}$CO(2$-$1) line corresponds more faithfully to
the optically thin case, and moreover the map extent coincides 
with that of the $^{12}$CO(1$-$0) map, so that a direct comparison 
is possible. The mismatch between the SEST beam size at this frequency and the adopted map spacing was fixed by assigning the integrated intensity at each position to the entire grid element, effectively replacing the smaller $^{13}$CO(2$-$1) beam size by the $^{12}$CO(1$-$0) beam size.
The obtained mass amounts to 
$M_{^{13}\mathrm{CO}(2-1)}=7.6 \times 10^3$~$M_{\odot}$, which is compatible with that obtained by means of the previous empirical method.
The remaining discrepancy can be explained by taking into account that the number of usable $^{13}$CO(2$-$1) spectra, contributing to the total map mass estimate, is markedly lower than in the case of $^{12}$CO(1$-$0), due to lower $S/N$, and taking into account that the LTE approximation method seems to typically underestimate the $^{13}$CO true column densities \citep{pad00}. Finally, we should note that these two lines, $^{12}$CO(1$-$0) and $^{13}$CO(2$-$1), map different parts of the cloud because of their different optical depths, and also that, evidently, the empirical method based on the $^{12}$CO(1$-$0) emission is quite
reliable in a statistical way, but for individual clouds it may
give inaccurate results. Therefore, given these considerarions, the two mass estimates
derived here can be considered to be in good agreement.

The column density of $^{13}$CO, derived from $^{13}$CO(2$-$1), at the centre position can also be used to estimate the optical extinction $A_V$ towards the cluster by applying a linear relation: 
\begin{equation}
A_V=c_1~N(^{13}{\rm CO})+c_2
\end{equation}

\noindent
Several estimates of the $c_1$ and $c_2$ 
parameters can be found in the literature, varying with the observed region 
and the calibration technique. Here, we adopt the results of 
\citet{fre82} for the Taurus region, $c_1=7.1 \times 10^{-16}$ mag~cm$^{-2}$ and $c_2=1.0$~mag respectively.
Thus, the obtained visual extinction estimate
is $A_V=2.3$~mag, much smaller than the value obtained from the $^{12}$CO(1$-$0) data.

Repeating the same procedure, now for the $^{13}$CO(1$-$0) line, but with the
aforementioned caveats about optical thickness and map size, and about the
accuracy of the LTE technique, we obtain a total mass $M_{^{13}\mathrm{CO(1-0)}}=8.6 \times 10^3$~$M_{\odot}$, and a visual extinction of $A_V=3.1$~mag toward the cluster line of sight.

\subsubsection{Structure of the gas emission}

Once derived the global characteristics of the region mapped in CO, we focus on the structure of the gas emission, analysing
in particular the cloud component that appears to be spatially associated 
with the IRAS~07527-3446 cluster. This can be done by applying 
a cloud decomposition algorithm that analyses the R.A.--Dec.--V$_{\rm LSR}$ data cube, assigning channels of observed spectra to different ``clumps'', and estimates the masses of these clumps separately.
A typical choice is the CLUMPFIND code \citep{wil94}; even though some 
problems can arise when applying this algorithm to very complex velocity fields
\citep[strong dependence from the initial setup, see e.g., discussion in][]{eli07},
in this case the input data cube is such that the cloud decomposition turned
out to be stable. Because of the higher $S/N$ in the spectra, the $^{12}$CO(1$-$0) data
cube was chosen for decomposition. The input data requested  
by the algorithm are the radiation temperature threshold to start the search for clumps, and the level increment. These were defined to be: $T_{\mathrm{min}}=0.5$~K, and $\Delta T=0.5$~K, in both cases more than two times higher than the rms noise of the line, as suggested by \citet{wil94}.

Four clumps were found, and in particular one of them (hereafter 
\textit{clump A}) is centred on the strongest peak of the map, i.e. the 
cluster position. The output average radius provided by CLUMPFIND is drawn in Fig.~\ref{map12}, as a dashed circle that encloses the entire 
area imaged in NIR. The properties of all four clumps are 
summarized in Table~\ref{clumptab}. In positions where they spatially overlap, spectral channels are attributed to different velocity components and then assigned to different clumps.  Because of the small extent of the map, all the clumps are
flagged by the algorithm as containing pixels that lie on the
edge of the investigated area and are expected to
extend outside it as well, and their average size is evaluated in
this perspective. Clumps B, C, and, mainly, D are more affected
by this uncertainty.
Applying to clump A the empirical method as above, we a mass estimated $M_A$ to be $3.3 \times 10^3$~$M_{\odot}$. This value can be considered as the mass of the gas component presently associated with the cluster, as obtained directly from $^{12}$CO(1$-$0) brightness.

\begin{table*}[ht]
\caption{Properties of $^{12}$CO(1$-$0) detected clumps \label{clumptab}}
\begin{tabular}{lccccccc}
\hline\hline
Clump  & Map offset          & Diameter  & $V_{\mathrm{LSR}}$  &  $\Delta V$ &  $M_{\mathrm{^{12}CO}}$ & $M_{\mathrm{vir}}$  \\
       & ($\arcsec,\arcsec$) &     (pc)      & (km s$^{-1}$) & (km s$^{-1}$)     & (M$_{\odot}$)          & (M$_{\odot}$)      \\
\hline
A      & (0,0)               & 18.7          & 89.4        & 1.24         & 3.3 10$^3$             & 2.7 10$^3$       \\
B      & (-46,+132)          & 19.1          & 91.4        & 1.68         & 4.0 10$^3$             & 5.2 10$^3$       \\
C      & (-132,-92)          & 16.2          & 89.2        & 0.99         & 2.1 10$^3$             & 1.5 10$^3$       \\
D      & (-230,-132)         & 15.7          & 87.8        & 0.81         & 0.8 10$^3$             & 1.0 10$^2$       \\
\hline
\end{tabular}\end{table*} 

Using the average velocity dispersion provided by the output of CLUMPFIND, 
and combining this with the estimated average radius, the virial mass for these clumps is estimated and given by:
\begin{equation}
M_{\mathrm{vir}}=210\left(\frac{R}{\mathrm{pc}} \right)
\left(\frac{\Delta V}{\mathrm{km~s^{-1}}} \right)^2  \qquad (\mathit{M_{\odot}})\;,
\end{equation}
where the multiplicative constant corresponds to the case of a
uniform clump density (for density profiles varying as $\rho\propto r^{-1}$
or as $\rho\propto r^{-2}$, the constant is replaced by 190
and 126, respectively). Virial masses of the four identified clumps
are listed in Table~\ref{clumptab}. For all these clumps, the ratio
of virial masses to those obtained from the $^{12}$CO(1$-$0)
integrated intensity is within 20--30\% of unity. This situation
is then compatible with the virial equilibrium; however, the uncertainties in
these mass estimates are too large to allow a robust conclusion.

Finally, the excitation temperature $T_{\rm ex}$ can be derived from the line intensity ratio of a molecular isotope, under the assumption of optically thin and LTE regime. For the two $^{13}$CO lines at the (0,0) position of the map, the line ratio is $\sim 1$ (see also Fig.~\ref{spectra}, panel $a$) and we derive an excitation temperature $T_{\rm ex}=7.7$~K. On the other hand, as written above, in the LTE assumption it is also possible to derive the excitation temperature from the thick line of $^{12}$CO(1$-$0), and in this point we obtain $T_{\rm ex}=12.5$~K . Both these values are consistent with the range of 6-15~K found by \citet{brand95} for far outer Galaxy clouds, for which these authors do not note significant differences with those present in the inner Galaxy part. On the other hand, \citet{mead88} proposed a different scenario, with a temperature gradient and lower temperatures in the outer Galaxy, being able to reproduce their data with a value of 7~K, not significantly different from that we obtain from the ratio of the $^{13}$CO lines. 

\subsection{The stellar content}

\subsubsection{Reddening and cluster membership}

\begin{figure}
   \centering
   \includegraphics[width=9cm]{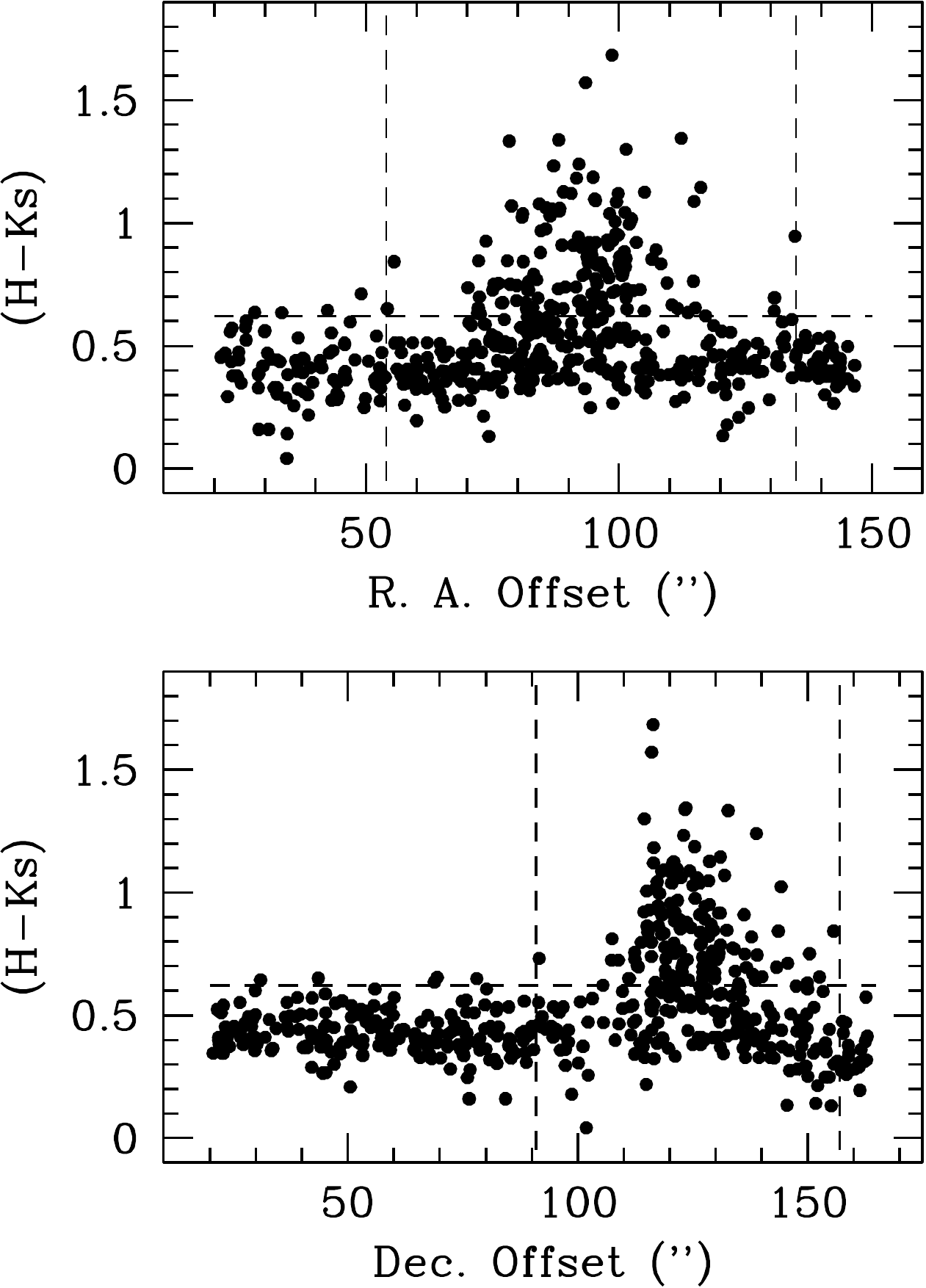}
   \caption{
{\it a (top):} Plot of the observed $(H-K_S)$ colour as a function of the R.~A. offset relative to the lower left (SE corner) of the upper panel of Fig.~\ref{K}. 
The reddening effect ($(H-K_S\geq 0.62$, horizontal dashed line indicating the field star average plus $2\sigma$) of a dense cloud core is clearly seen for values of R.~A. offsets between approximately 54$^{\prime\prime}$ and 135$^{\prime\prime}$ (indicated by vertical dashed lines).
{\it b (bottom):} The same as {\it a} but as a function of Dec. offset. The presence of the dense cloud core occurs for Dec. offsets between approximately 91$^{\prime\prime}$ and 157$^{\prime\prime}$.
	} 
	\label{hkxy}%
    \end{figure}

The top panel of Fig.~\ref{hkxy} shows the spatial distribution of the $(H-K_S)$ colour across the image, as a function of the R.~A. offset relative to the lower left (SE corner) of Fig.~\ref{K}. The bottom panel is the same but as a function of Dec. offset. The redder values of $(H-K_S)$ seen for values of R.A. offset between 54\arcsec\ and 135\arcsec\ and Dec. offset between 91\arcsec\ and 157\arcsec\ indicate that a large number of sources in this region exhibit colour excesses with respect to field stars. These ranges of coordinates define a box that we refer to as the ``cluster region'' (shown as the dashed box in Fig.~\ref{K}). In the cluster region, there is an enhancement in the visual extinction traced by redder (H-Ks) colours, corresponding to the presence of the cluster parental molecular cloud. This signals the embedded nature and youth of the cluster sources.

%

\begin{figure}
   \centering
   \includegraphics[width=9cm]{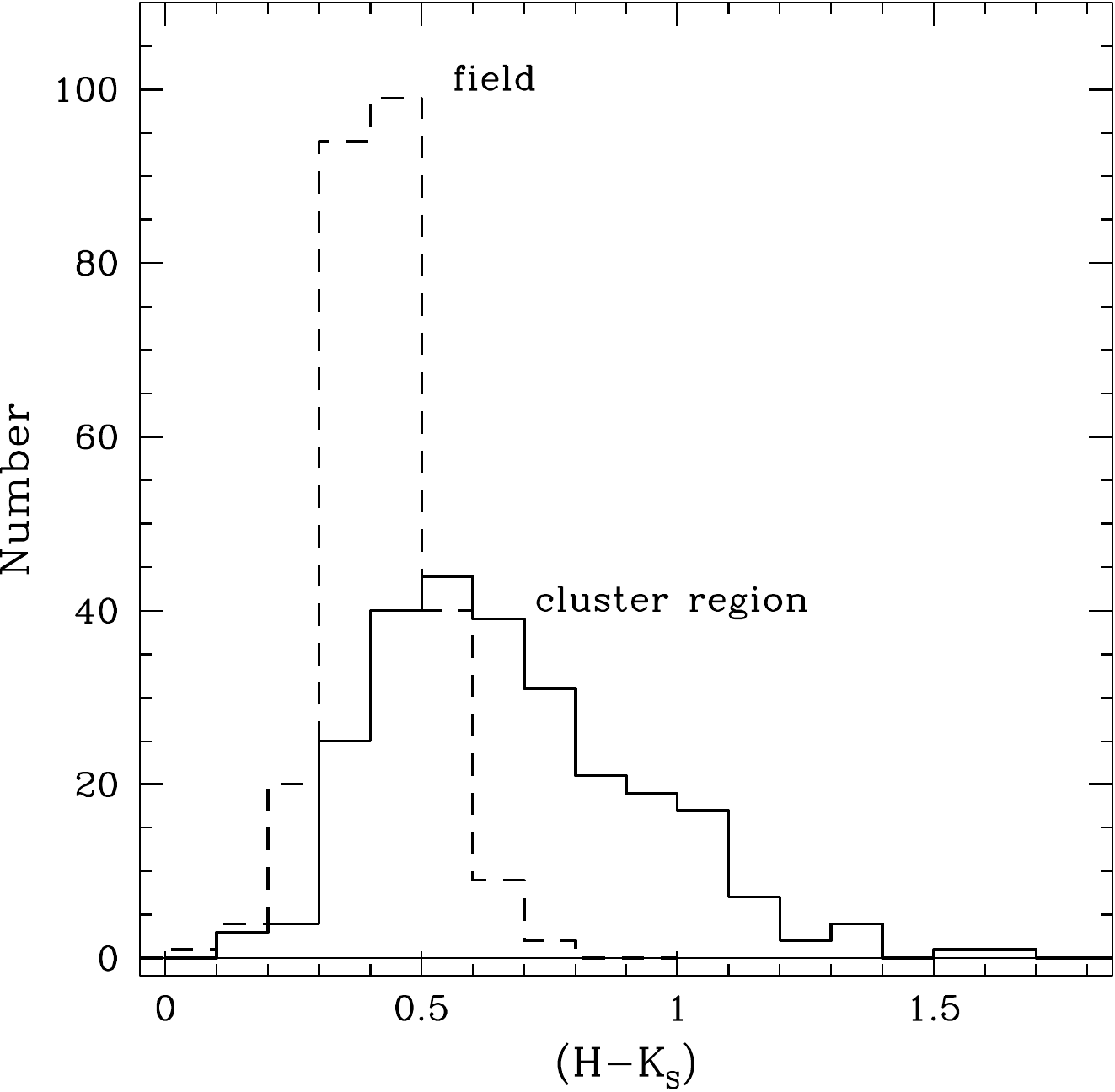}
   \caption{
Histogram of the observed $(H-K_S)$ colours for sources inside (solid line) or outside (dashed line) the cluster region.
	} 
	\label{two_histos}%
    \end{figure}

We plot in Fig.~\ref{two_histos} the histograms of the $(H-K_S)$ colours of the sources detected in both the $H$ and $K_S$-band images.
Sources closer than about 20\arcsec\ to the left or the lower borders were excluded. 
The solid line represents the histogram of 258 sources located \textit{inside} the cluster region. The mean value of their $(H-K_S)$ colours is 0.67 and the standard deviation is 0.26.
The dashed line corresponds to the histogram of the 269 sources located \textit{outside} the cluster region (contained mostly in the southern part of the image). Here, the sources, mostly foreground field sources, exhibit a close to symmetrical histogram and have very different statistical values: their mean value is $(H-K_S)_{\rm field}= 0.42$, with a standard deviation of 0.10. 
These values were used to draw the horizontal line in Fig.~\ref{hkxy}, which is placed at a value of $(H-K_S)=0.62$, i.e., at a value of the mean of $(H-K_S)_{\rm field}$ plus $2\sigma$.

%

As argued in the next paragraphs, we consider a sufficient condition for cluster membership that a star is located inside the cluster region {\it and} has a colour of $(H-K)\ge 0.62$.

Along the horizontal axis of Fig.~\ref{hkxy}, for all values of R.~A., there are values of $(H-K_S)$ between about 0.05 and 0.62. The reddening effect of the cloud clearly stands out as enhanced values of $(H-K_S)$ for values of R.~A. offset between approximately 54$^{\prime\prime}$ and 135$^{\prime\prime}$ (a range of 81$^{\prime\prime}$ -- about 4 pc at the source distance of 10.3 kpc -- indicated by the two vertical dashed lines). 
A similar conclusion is reached by analysis of the values of $(H-K_S)$ as a function of Dec. offset.
For the general case, this colour excess of $(H-K_S)=0.62$ would just separate the objects into two groups: a group of foreground stars and a group of \{embedded + background\} stars. However, given that at this distance and location in the Galaxy, there are very few background stars,
this reddening effect
separates foreground from embedded stars and thus effectively selects cluster members.

Consequently, we take the diagrams of Fig.~\ref{hkxy} and the association of CO emission 
as evidence of a cloud core and we use it to define the ``cluster region'' as this rectangular region of about $81^{\prime\prime} \times 66^{\prime\prime}$ on each side. The cluster region is shown in the close-up panel of Fig.~\ref{K}. A star that is located in the cluster region {\it and\/} has $(H-K_S)>0.62$ is considered here to be a cluster member. Evidently, this is a sufficient but not a necessary condition. There may be additional cluster members outside the cluster region or inside it with bluer values of $(H-K_S)$. One hundred and thirty-six sources verify these criteria and are thus good candidate cluster members. This number is in good agreement with the expected number of 138 cluster members derived from the surface density plot in Section~3.1. It represents a lower limit to the number of young stellar cluster members detected in the $H$ and $K_S$-band images. Photometry of these sources is given in Table~2\footnote{Table~2 is available in electronic form at the Centre de Donn\'ees Astronomiques de Strasbourg (CDS) via anonymous ftp to cdsarc.u-strasbg.fr (130.79.128.5) or via http://cdsweb.u-strasbg.fr/cgi-bin/qcat?J/A+A/}.
In this table, Cols. (1), (2), and (3) identify the stars giving, respectively, the running ID number, and the equatorial coordinates (R.A. and Dec., epoch 2000); Columns (4), (5), and (6) give the photometric indices $K_S$, $(H-K_S)$, and $(J-K_S)$.

\subsubsection{Young stellar objects}

\begin{figure}
   \centering
   \includegraphics[width=9cm]{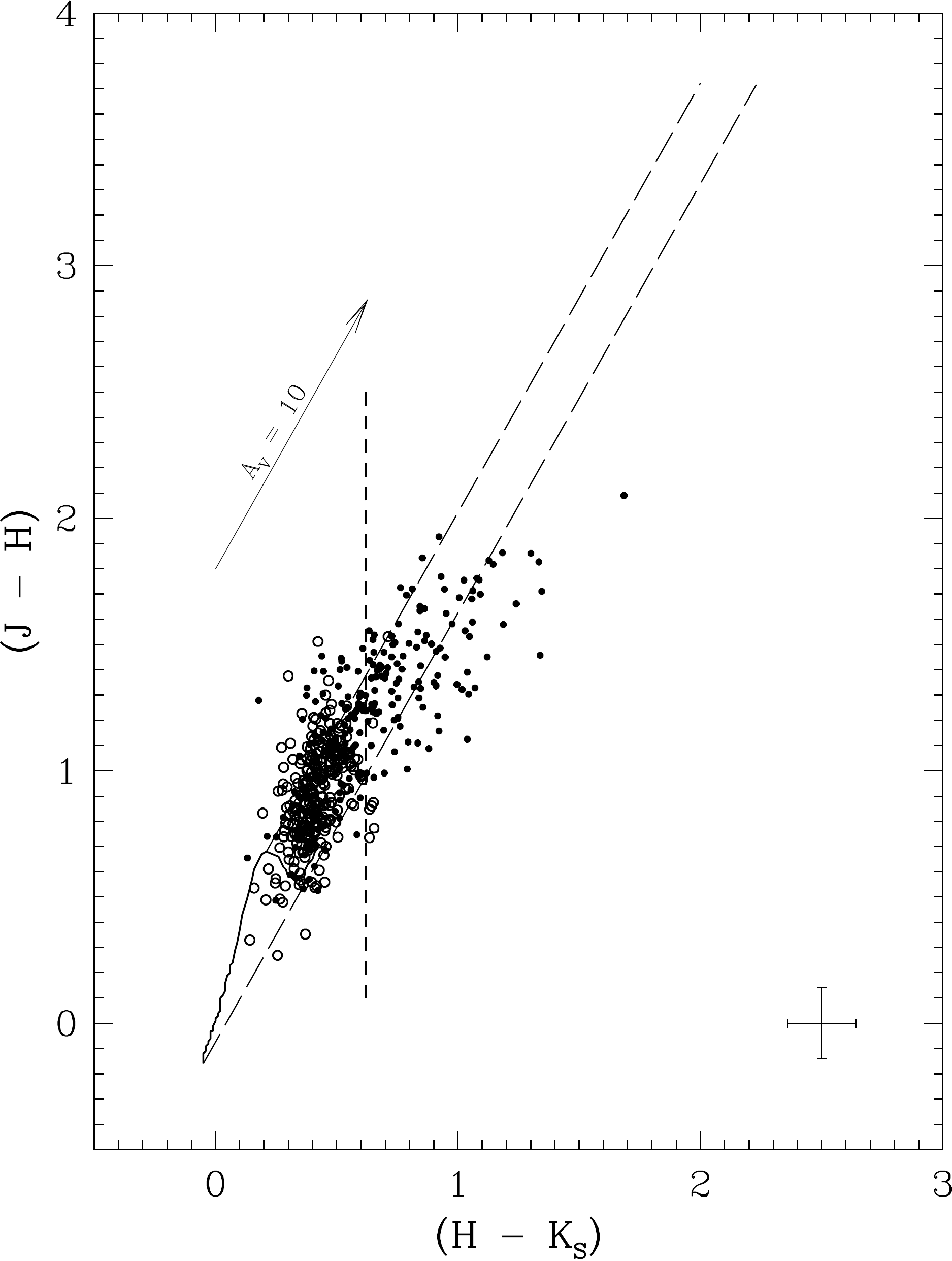}
   \caption{
Near-infrared colour-colour diagram of the region towards IRAS~07527-3446 seen in Fig.~\ref{K}. 
Filled circles represent the sources contained in the ``cluster region'' (see text). 
Open circles are sources outside the cluster region.
The solid line represents the loci of unreddened main-sequence stars \citep{bessel88} while long-dash lines indicate the reddening band. The reddening vector indicates the direction of the shift produced by extinction by dust with standard properties \citep{rieke85}. 
The location of the vertical dashed line, derived from Fig.\ref{hkxy}, is at $(H-K_S) = 0.62$.  Notice that virtually all the sources to the right of this line are in the cluster region. The cross in the lower right corner represents a typical error bar.
	} 
	\label{cc}%
    \end{figure}

In the region shown in the top panel of Fig.~\ref{K}, 
using the 500 point sources detected in all three $J$, $H$, and $K_S$-bands,  
we have plotted the near-infrared colour-colour diagram, $(J-H)$ versus $(H-K_S)$. This is shown in Fig.~\ref{cc}.
In this plot, filled circles represent 
sources contained in the $81''\times 66''$ central region (``cluster region'', where most sources are likely to be members of the cluster). Open circles are sources outside this region.
The location of the vertical dashed line, derived from Fig.~\ref{hkxy}, is at $(H-K_S) = 0.62$. Most of the sources to the right of this line ($(H-K_S) > 0.62$) are  in the cluster region.
Sources to the left of the dotted line are both inside and outside the cluster region.

%

The solid line represents the loci of unreddened main-sequence stars \citep{bessel88}. Most stars are located within the reddening band (long-dash lines), where stars appear if they are main-sequence stars reddened according to the interstellar extinction law \citep{rieke85}, which defines the reddening vector (traced here for $A_V=10$).
Pre-main-sequence YSOs that have had time to clear the inner regions of their circumstellar discs lie in this region as well. Giant stars appear slightly above this band. Stars located to the right of the reddening band are likely to be embedded young star objects with infrared excess emission due to the presence of circumstellar material \citep{adams87}. 

We note that not many sources lie in the infrared excess region. Among the sources with values $(H-K_S) \ge 0.62$, the fraction of sources in the infrared excess region is 40\%. This relatively low number of sources in this region suggests that most sources ($60$\%) have dissipated the inner parts of their circumstellar discs that are responsible for the $(H-K_S)$ excess. This result indicates that the age of this cluster is likely to be about 3-5 Myr. This is because inner disc frequencies in young clusters have been found to decrease steadily as the cluster ages approach about 6 Myr \citep{haisch01}, with about 50\% of the sources exhibiting excess emission from circumstellar discs at 3 Myr and very few sources at ages older than about 6 Myr. The cluster could be younger if it contains Class~II sources with circumstellar discs that cannot be detected in a JHK colour-colour diagram \citep{lada95}.

For the sources that lie inside the reddening band, the highest value of $(H-K_S)$ is about 1.08. Using the mean value of $(H-K_S)=0.42$ for field stars (according to Fig.~\ref{two_histos}), we obtain a colour excess $E(H-K_S) = 0.66$ due to intra-cloud extinction, which corresponds to about a maximum visual extinction produced by the cloud core of about $A_V=11$ \citep{rieke85}.

\begin{figure}
   \centering
   \includegraphics[width=9cm]{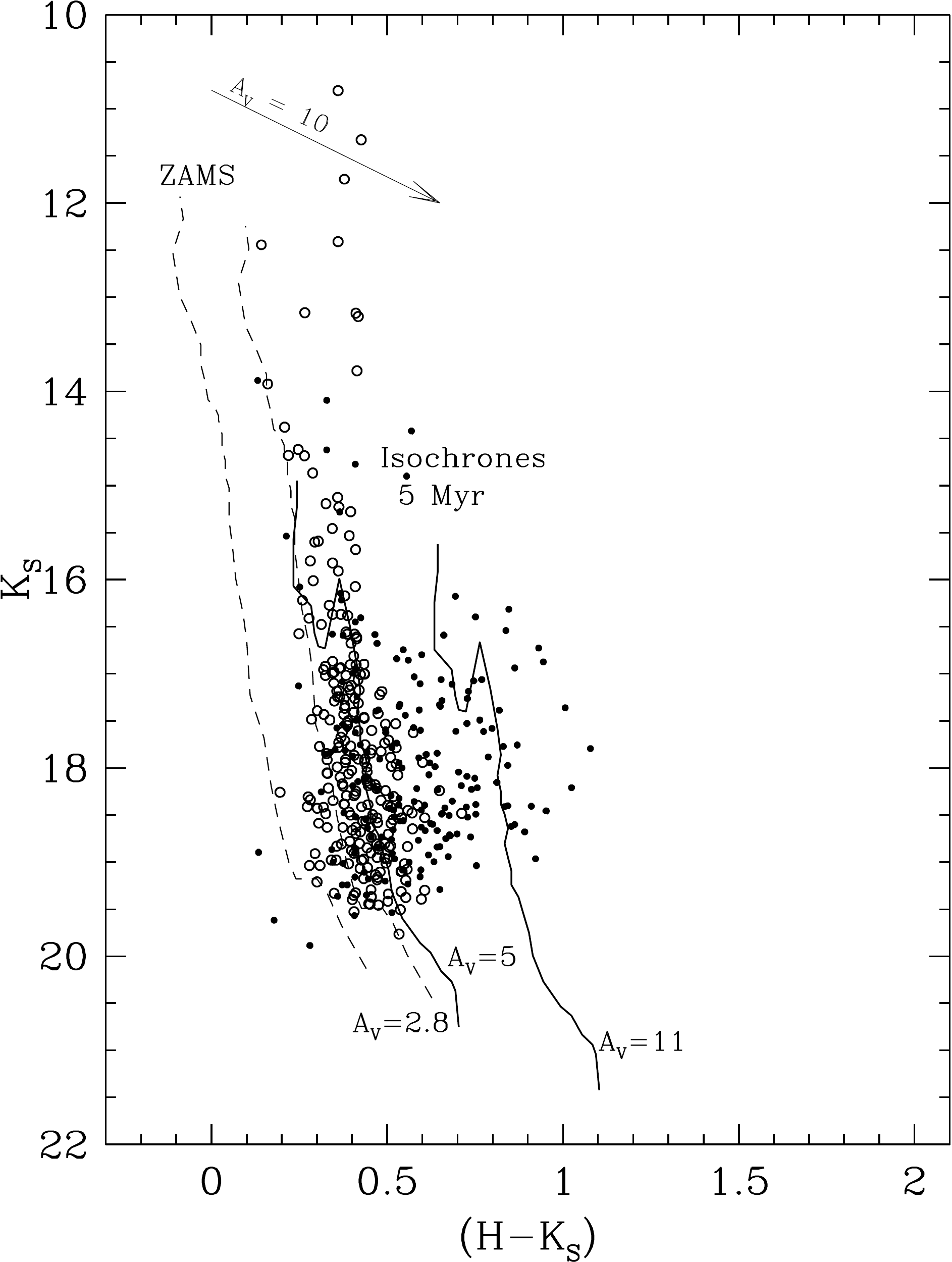}
   \caption{
Colour-magnitude diagram of the region towards IRAS~07527-3446 seen in Fig.~\ref{K}.
Filled circles and open circles have the same meaning as in Fig.~11 but excluding sources with circumstellar emission. The two dashed lines are, respectively, the zero-age main-sequence \citep[ZAMS;][]{siess00} for an arbitrary distance with zero extinction and with $A_V=2.8$. The two solid lines are the 5 Myr isochrone with extinctions $A_V=5$ and $A_V=11$, respectively. The reddening vector indicates the direction of the shift produced by extinction by dust with standard properties \citep{rieke85}.
 	} 
	\label{isochrones}%
    \end{figure}

An additional estimate of the age (or range of ages) for this cluster can be obtained from the colour-magnitude diagram, $K_S$ vs. $(H-K_S)$, plotted in Fig.~\ref{isochrones}, where open circles have the same meaning as in Fig.~\ref{cc}, and filled circles are as in Fig.~\ref{cc} but exclude sources with circumstellar emission (outside the reddening band). The two dashed lines are the zero-age main-sequence (ZAMS) for an arbitrary distance with zero extinction and with $A_V=2.8$, respectively. This extincted ZAMS represents a rough fit to the open circles (which are spread mostly along a nearly vertical band). The several brightest sources, which are clearly blue, are field giants (non-cluster members).
The large spread in the open circles around the fit is expected since the foreground stars are located at variable distances and their different lines-of-sight may be affected by variable values of ISM extinction. Thus,
this value of extinction ($A_V=2.8$) represents a rough lower limit to the extinction in the region: it is caused by the foreground diffuse ISM, and does not include the internal cloud extinction.

%

Focusing our attention now on the assumed cluster members (filled circles), the location of a source on this colour-magnitude diagram is the result of a combination of cluster distance, cluster age, and extinction along the line-of-sight towards the source. 
Assuming a fixed distance for the cluster members, vertical spread (scatter in magnitude $K_S$) can be caused mostly by a spread in age. Variable extinction will only produce a small vertical dispersion because the extinction vector is almost horizontal (see Fig.~\ref{isochrones}).
On the other hand, a horizontal spread in this diagram (scatter in colour) can be due to different ages, variable spatial extinction, or a combination of both. 
The existence of variable extinction, with highly different amounts of dust across young clusters is expected. Non-coeval star-formation cannot be excluded either. Thus, we have tried to fit the filled circles using pre-main-sequence isochrones reddened by variable amounts of extinction. 

Fitting pre-main-sequence isochrones to young stellar clusters is a rather uncertain process but it can give a rough estimate of the age or a range of ages of a young cluster.  We have used the kinematic distance of 10.3 kpc and the set of \citet{siess00} isochrones with masses ranging from 0.1 to 7 $M_{\odot}$. 
Given the absence of metallicity information, we have adopted solar metallicity. A set of isochrones of different ages and extinctions were generated and plotted in the colour-magnitude diagram. Figure~\ref{isochrones} shows a possible fit to the filled circles: an isochrone of 5 Myr with extinctions varying from $A_V=5$ to $A_V=11$, in good agreement with the values of $A_V$ derived from the $^{12}$CO gas column densities. The uncertainty in the isochrone models imply an age spread of about 2~Myr.
The value of $A_V\sim 5$ for the least-extincted cluster sources is in good agreement with the value of $(H-K_S)=0.62$ for the same sources. Using an intrinsic (non-reddened) average $(H-K_S)$ colour of about 0.25 for field stars \citep[mostly main-sequence and red giants;][]{koornneef83,bessel88}, the colour excess $E(H-K_S)=0.62-0.25=0.37$ corresponds to an $A_V= 15.3 \times E(H-K_S) = 5.6$ \citep{rieke85}.
Isochrones of younger ages will fit only the lower mass-end (the fainter sources). They would require the absence of any intermediate-mass stars restricting the cluster population to low-mass stars only. Assuming an age of 5 Myr for this cluster, and considering the faintest cluster member, the lowest mass detected in our $K_S$-band image is about 0.4 $M_{\odot}$ for $A_V=5$, and about 0.8 $M_{\odot}$ for $A_V=11$.
Similarly, the highest-mass star present in our image is estimated to be between 5 and 9 $M_{\odot}$. This value is in good agreement with the absence of massive stars inferred from the lack of VLA continuum emission. 

According to the VLA 6~cm observations (Sect.~2.3), and down to the sensitivity of these observations, no radio-continuum emission was detected at this wavelength. Setting the lower-limit detection for the 6~cm continuum emission to be at 4$\sigma$,
and adopting a distance of 10.3~kpc, the rate of ionizing photons required to maintain the ionization of a possible HII region up to that level is $\dot{N_i} \simeq ~6.7 \times 10^{45}$~s$^{-1}$ (under the assumption of an optically thin, homogeneous spherical HII region with constant temperature of $T=10^{4}$ K; e.g., Rodr\'iguez et al. 1980). This implies an upper limit of $\sim 6 \times 10^3$~$L_\odot$ \citep[][corresponding to a B1 zero-age main-sequence star with M~$\sim$~12~$M_{\odot}$]{panagia73} for the luminosity of the individual members of the IRAS~07527-3446 cluster.

Furthermore, the luminosity of the cluster region is dominated by the mid and far-infrared flux as measured by IRAS. We estimate this $L_{\rm FIR}$ to be about $3.5\times 10^3 L_{\odot}$. Even assuming (unrealistically) that all this luminosity is produced by a single star, this would set an upper limit of about 8 $M_{\odot}$ for any massive star present in this cluster, again in good agreement with the above estimates.

\subsubsection{Luminosity function}

   \begin{figure}
   \centering
   \includegraphics[width=9cm]{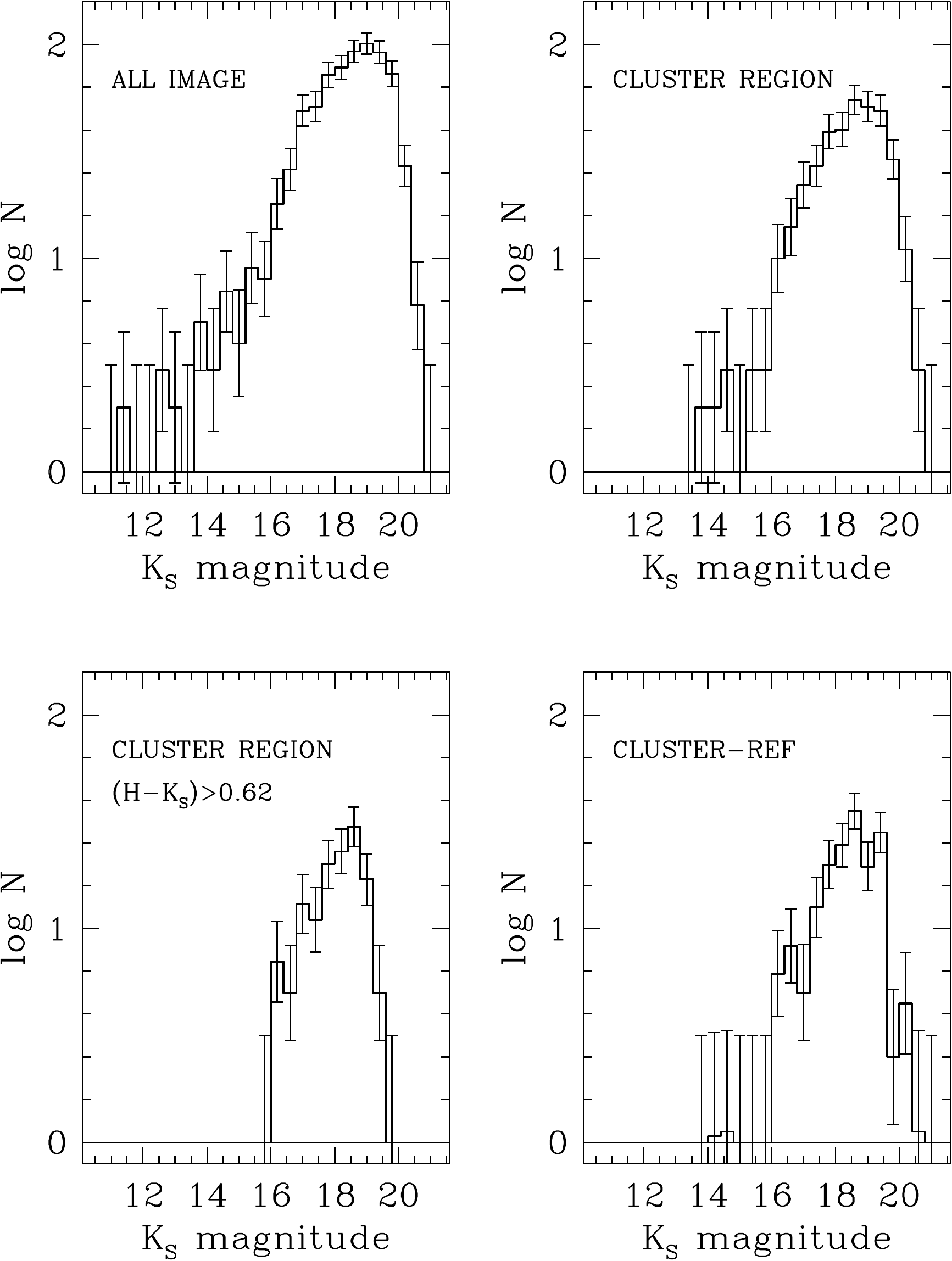}
   \caption{
Plots of the logarithm of the number of stars as a function of $K_S$. The upper left panel is for all the stars in the $K_S$-band image. The upper right panel is for the stars contained in the cluster region. 
The two lower panels are obtained from this panel: {\it i)} by restricting to the stars in the cluster region having $(H-K_S) > 0.62$ (lower left panel); {\it ii)} by subtracting the stars in a reference region (lower right panel).
Error bars are 1 $\sigma$ counting statistics.
} 
	\label{histo}%
    \end{figure}
%

The $K_S$-band luminosity function of this source was constructed using the results of our $K_S$-band photometry. The cluster members were grouped into bins of 0.4 mag width. Figure~\ref{histo} shows the results. The upper left panel gives the histogram of $K_S$ magnitude for all sources in the image, whereas the upper right panel is the histogram for all sources located inside the cluster region.
In an attempt to correct for foreground field stars, we present in the lower left panel the histogram of the reddest sources inside the cluster region, which should be composed of essentially cluster members. No attempt has been made to deredden the sources. 
A different attempt to correct for the foreground stars consisted of considering the southern part of the image (where few or no cluster members exist) as a reference field. The histogram of the reference field (not shown) was scaled to the area of the cluster region and subtracted from the histogram of the cluster region yielding the lower right panel.

We consider the straight line ($\log N = a + b K_S $) that best fits each histogram in the range of bins, where there are statistically significant numbers of sources ($K_S \sim 16-19$).
Interestingly, the slopes $a$ of the four histograms, in that range of bins, are very similar: $0.32\pm 0.03$, $0.31\pm 0.03$, $0.31\pm 0.06$, and $0.33\pm 0.07$, respectively.
These values are similar to those found for nearby clusters which range from 0.30 to 0.40 \citep[e.g.,][]{devine08}. Therefore, no statistical difference could be found in the $K_S$-band luminosity function of this distant cluster located in the far outer Galaxy.

The slope found for this cluster seems to rule out a continuous star-formation scenario, favouring instead a coeval star-formation case \citep{lada95}. In this case, the KLF has also been associated with cluster age, with KLFs of young clusters becoming shallower as they become older. The value of about 0.32 is compatible with an cluster age of about 4 Myr.

It is common to use the $K_S$-band luminosity function to obtain the initial mass function IMF of a cluster. We have not attempted to do so because that conversion requires the knowledge of several factors involved, such as the pre-main-sequence mass-luminosity relationship, the population incompleteness, the extinction, the cluster age, and a proper background-source subtraction \citep{muench00}. None of
these parameters is well known and would introduce large uncertainties into an IMF calculation.

Bearing in mind that we do not estimate any individual masses, but taking into consideration the above conclusions, we can make a rough estimate of the total mass present in the stellar content of this cluster. We assume that we have detected all stars down to 0.4~$M_\odot$ and we use a Salpeter IMF down to 1~$M_\odot$. Below 1~$M_\odot$, we use a shallower slope \citep[of $-1.2$;][]{stahler04}. We obtain a total stellar mass of about 200~$M_\odot$ inside the cluster region. 
Furthermore, assuming that clump A is the one that gave origin to the young stellar cluster, using the mass of this clump as the mass of the gas, we obtain a star-formation efficiency of $M_{\mathrm stars} / (M_{\mathrm stars} + M_A) \simeq $~5.7\%. Even though the uncertainty is large, these values are similar to other values of star-formation efficiencies found in cluster environments within the local star-formation regions \citep[e.g., L1630,][]{lada99}.

\bigskip

The presence and properties of this molecular cloud containing this young stellar cluster at this location in the far outer Galactic disc constitutes further evidence that the properties of the far outer Galaxy molecular clouds and their star-formation activity are quite similar to those of molecular clouds in the inner Galaxy. This is also true for the few previous cases where we have found clusters in the far outer Galaxy: IRAS~07255-2012 \citep{santos00}, and IRAS~06361$-$0142 \citep{yun07}. Another common feature for these 3 clusters, all located in the far outer Galaxy, is the absence of very massive (O and B) stars. Albeit the small number statistics, there seems to be a tendency for distant far outer Galaxy molecular clouds not to form very massive stars.

\section{Summary}

We have discovered a new young stellar cluster seen towards a molecular cloud located at a distance of about 10 kpc and a 
Galactocentric distance of about 15 kpc, well within the far outer Galaxy.

The cluster (with at least 136 objects) 
is detected in our near-infrared images as a strong enhancement in the surface stellar density of our near-IR images. The cluster is contained in a region of about 4.0 $\times$ 3.3 pc$^2$ centred close to IRAS 07527--3446. It appears to consist of low and intermediate-mass young reddened stars with a large fraction having cleared the inner regions of their circumstellar discs. The observations are compatible with a $\le$ 5~Myr cluster with variable spatial extinction between $A_V=5$ and $A_V=11$. 
The slope of the $K_S$-band luminosity function is estimated to be 0.33, similar to those of young clusters in nearby star-forming regions and compatible with a coeval star-formation scenario.

The molecular cloud, as traced by CO($J$=1$-$0), extends over a region of $\sim 20 \times 15$~pc$^2$. Our estimates of the mass of the molecular gas in this region, range from a few to several tens of thousands of solar masses, depending
on the adopted technique and on the line transition used.
In particular, decomposing the CO emission into clumps, we find that a clump is clearly associated with the cluster position, and has a mass of $3.3 \times 10^3$ M$_\odot$. The average velocity dispersion of this clump suggests that it is not in virial equilibrium. In addition, the extinction $A_V$ towards the cluster position, derived by means of different column-density methods, was found to be in the range $A_V \simeq 3 - 8$~mag.
We estimate a value of $\sim$~6\% for the star-formation efficiency of this molecular region.

These findings confirm previous results that the distant outer Galaxy continues to be active in the production of new and rich stellar clusters, 
with the physical conditions required for the formation of rich clusters continuing to be met in the very distant environment of the outer Galactic disc.

\begin{acknowledgements}
This work has been partly supported by the Portuguese Funda\c{c}\~ao
para a Ci\^encia e Tecnologia (FCT) and by the European Commission FP6 Marie Curie Research Training Network ``CONSTELLATION'' (MRTN-CT-2006-035890).
This research made use of the NASA/ IPAC Infrared Science Archive, which is operated by the Jet Propulsion Laboratory, California Institute of Technology, under contract with the National Aeronautics and Space Administration.
This research also made use of the
SIMBAD database, operated at CDS, Strasbourg, France, as well as SAOImage
DS9, developed by the Smithsonian Astrophysical Observatory.

\end{acknowledgements}

\end{document}